\documentclass[aps,amsfonts,amsmath,prd,preprint,nofootinbib]{revtex4}
\usepackage{epsf,mathrsfs}
\usepackage{graphicx} 

\newcommand{\beq}{\begin{equation}}
\newcommand{\eeq}{\end{equation}}
\newcommand{\bp}{{\bf P}}
\newcommand{\bk}{{\bf K}}

\begin{document}

\title{Living with ghosts in Lorentz invariant theories}

\author{Jaume Garriga$^{1}$, Alexander Vilenkin$^2$.}

\address{$^1$ Departament de F{\'\i}sica Fonamental i \\Institut de Ci{\`e}ncies del Cosmos, 
Universitat de Barcelona,\\
Mart{\'\i}\ i Franqu{\`e}s 1, 08028 Barcelona, Spain
\\
$^2$ Institute of Cosmology, Department of Physics and Astronomy,\\
Tufts University, Medford, MA 02155, USA}

\begin{abstract}

We  argue that theories with ghosts may have a long lived vacuum 
state even if all interactions are Lorentz preserving. In space-time dimension $D=2$,  we consider the tree level 
decay rate of the vacuum into ghosts and ordinary particles mediated by non-derivative interactions, showing that 
this is finite and logarithmically growing in time.  For $D>2$, the decay rate is divergent unless we assume that the 
interaction between ordinary matter and the ghost sector is soft in the UV, so that it can be described in terms of non-local 
form factors rather than point-like vertices. We provide an example of a nonlocal gravitational-strength
interaction between the two sectors, which appears to satisfy all observational constraints.
\end{abstract}

\maketitle

\section{Introduction} 

Ghosts have the reputation of being disastrous for particle theories.   Any theory that allows ghosts, that is, particles with negative 
kinetic energy, to appear in physical states suffers from an instability with respect to spontaneous creation of ghosts and ordinary particles 
with zero total energy and momentum.  This process may be strongly suppressed if the coupling between the particle and ghost sectors is very weak.  
But even in the absence of any direct coupling both sectors must be coupled to gravity, and so gravity-mediated vacuum decay cannot be avoided.

Furthermore, the following standard argument \cite{Carroll,Cline,Sundrum} indicates that the corresponding decay rate is divergent.  
Each particle-ghost nucleation event can be characterized by the 4-momentum $P^\mu$ of the created particles, the 4-momentum of ghosts being 
$K^\mu=-P^\mu$.  The vacuum decay rate per unit spacetime volume can depend on $P^{\mu}$ only through the invariant
\beq
s=P_\mu P^\mu.  
\label{s}
\eeq
Hence, the total rate will be given by an integral over $s$ and over all possible locations of the 4-vector $P^\mu$ on the mass shell (\ref{s}),
\beq
\Gamma = \int ds F(s) \int\frac{d^3 {\bf P}}{\sqrt{{\bf P}^2+s}},
\label{Gamma}
\eeq
where $F(s)$ is a model-dependent function.  The ${\bf P}$-integration is over the Lorentz group.  
This integral is manifestly divergent, which indicates that the theory is inconsistent.

On the other hand, ghosts have been proposed for some interesting cosmological roles.  Linde 
\cite{Linde} has argued that the smallness of the cosmological constant can be attributed to a ghost sector described 
by the same Lagrangian as the matter sector, but with opposite sign.  Kaplan \& Sundrum suggested \cite{Sundrum} that the 
existence of such a sector could follow from an approximate discrete symmetry they called "energy parity". This symmetry is 
broken by gravity, and a small cosmological constant is induced.  Theories with ghosts have also been suggested as models of 
dark energy \cite{darkenergy,Carroll,Cline,Antoniadis}.  These ideas provide motivation not to dismiss ghosts too lightly and to find a way to get around the vacuum decay 
catastrophe.  

A possible, if somewhat radical, way to do that was pointed out in Refs.~\cite{Cline,Sundrum}.  Assuming that the interaction between the particle and 
ghost sectors is mediated by gravity, they
suggested that both integrations in (\ref{Gamma}) are effectively cut off due to modifications in gravitational physics above some energy scale $\mu$.  
In order to induce a cutoff in the Lorentz group integral over ${\bf P}$, this new physics necessarily has to violate Lorentz invariance.  
With these assumptions, the vacuum decay rate can be estimated as 
\beq
\Gamma \sim \frac{\mu^8}{M_p^4},
\label{Gamma1}
\eeq
and it was argued in \cite{Cline} that for $\mu\ll 1$~MeV particle production at this rate out of vacuum is consistent with observations.  
The expected magnitude of the cosmological constant in this kind of models is 
\beq
\Lambda\sim \mu^4,
\eeq
which agrees with the observed value for $\mu\sim 10^{-3}$~eV.

Giving up Lorentz invariance is a high price to pay, and in the present paper we shall explore the possibility of living with ghosts in Lorentz 
invariant theories.  We shall argue that this is indeed possible.  The key point is that even if Lorentz invariance is not explicitly broken, 
a vacuum with ghosts can at best be metastable, so it could not have existed forever.  If this vacuum was created on some spacelike hypersurface 
$\Sigma$, this surface breaks Lorentz invariance spontaneously and can in principle introduce an effective cutoff for the ${\bf P}$-integration 
in (\ref{Gamma}).  
In a cosmological context, the ghost-carrying vacuum could exist inside certain kind of bubbles in models of eternal inflation.  
In this case, the role of $\Sigma$ would be played by the hypersurface where the phase transition occurs, from an earlier ghost-free vacuum
(or, if all vacua are ghost-carrying, then $\Sigma$ might be taken as the semiclassical beginning of the universe).
For simplicity, here we shall restrict attention to a Minkowski vacuum, and we shall assume that $\Sigma$ is a flat hypersurface.  
We shall see that in the presence of such a surface the boost integration in (\ref{Gamma}) is indeed regulated, 
provided that the interaction between the particle and ghost sectors which is responsible for the vacuum decay is non-local and has 
a finite characteristic time (and length) scale $\mu^{-1}$.  

Generic forms of non-locality can be problematic in quantum field theories, as they may lead to violations of causality.  Such violations, 
however, may be phenomenologically acceptable, as long as they are confined to sufficiently small scales.  This appears to be the case in 
a wide class of models \cite{joglekar1,joglecar2,Wise,Wise2}.
Another problem is that
the time ordered exponential in the standard perturbative expansion of the S-matrix is not well defined in a nonlocal theory.  Time ordering becomes 
frame-dependent, and as a result the Lorentz invariance of the S-matrix is not guaranteed.  Difficulties with the defininition of a meaningful  S-matrix for 
generic non-local interactions have been emphasized in \cite{Marnelius}. 

On the other hand, non-locality is acceptable in effective field theories, where it arises from integrating out degrees of freedom of 
an underlying causal theory. In particular, string theory is expected to be described by an effective field theory, where non-locality becomes 
important at energies comparable to the string scale.\footnote{Independent arguments for an intrinsic non-locality of gravity have been presented 
in the context of black hole evaporation and holography. See e.g. \cite{Giddings} and references therein.}  
We note also that even though a generic nonlocal theory may violate Lorentz invariance, there are some special theories which are known to be free 
of this problem; see, e.g., Refs.~\cite{Wise,Woodard}.  
We shall assume that the nonlocal interaction between particles and ghosts belongs to this special class.

Our basic idea can be illustrated by the following heuristic argument.
Suppose for simplicity that the ghost-infested vacuum is locally Minkowski and was created on a flat hypersurface $t=0$.  
Suppose further that in the frame where both particles and ghosts nucleate in their respective center of mass frames, 
that is, where ${\bf P}=0$, the spacetime region affected by the nucleation process is $\Delta t\sim |\Delta {\bf x}| \sim \mu^{-1}$.   
A nonzero value of ${\bf P}$ corresponds to the same process boosted to a frame moving with Lorentz factor $\gamma\sim |{\bf P}|/\mu$.  
For $\gamma\gg 1$, the affected region has the same spacetime volume, but is 
highly stretched in the direction of motion and in the time direction, e.g.,  
\beq
\Delta t\sim \gamma\mu^{-1}.  
\eeq
For vacuum decay occurring at time $t$, we should require $\Delta t \lesssim t$, since otherwise the nucleation region will stretch beyond the initial surface $t=0$.  The corresponding momentum cutoff is 
\beq
P_{max}\sim \mu^2 t.
\label{Pmax}
\eeq
The particle creation rate is then
\beq
\Gamma \sim \frac{\mu^8}{M_p^4}(\mu t)^2.
\label{Gamma2}
\eeq

Note that integration over the Lorentz group in Eq.~(\ref{Gamma}) comes from integration over external momenta and does not arise in loop diagrams.  
Hence, vacuum energy diagrams are cut off at $\mu$, not at $P_{max}$, and the gravitational contribution to the cosmological constant 
is still expected to be $\sim \mu^4$.

The possibility that the divergent vacuum decay rate may be regularized by a finite age of the universe has been briefly 
considered by Cline et al \cite{Cline} \footnote{ A related idea was recently discussed in footnote 5 of 
Ref. \cite{safety} (see also \cite{GCS}).}. They suggested that the resulting decay rate would be given by an expression 
similar to Eq.~(\ref{Gamma2}), where $\mu$ has the meaning of the high-energy cutoff of the theory.  
The main difference with our results is that the authors of \cite{Cline} did not recognize the essential role of 
non-locality in regulating the vacuum decay catastrophe. \footnote{ Starting from a local field theory 
with ghosts, the interaction between ordinary particles and ghosts can of course be thought of as non-local after 
the graviton (or any other field mediating the interaction between the two sectors) is integrated out. However, 
as we shall see in Section III, this will not make the rate of vacuum decay into the remaining species finite, 
even if the momentum squared $s_0$ of the mediator is bounded by a cut-off 
$s_0\lesssim \mu^2$. Moreover, since the graviton can also be present in external legs, nothing prevents a 
catastrophic decay rate of the vacuum into ghosts and gravitons through their local interaction.}

In the rest of this paper, we will study the decay rate of a vacuum with ghosts by a direct computation, showing agreement with the heuristic 
estimate (\ref{Gamma2}). The paper is organized as follows. In Section II we discuss the decay of a ghost-infested vacuum assuming a local interaction between ordinary particles and ghosts in $D$ spacetime dimensions.  We find that a finite vacuum lifetime does regulate the decay rate in $D=2$, but not in any higher number of dimensions.

In Section III, focussing now on the physically interesting case of $D=4$, we consider a simplified model with restricted non-locality. In this model,
interactions are local within each sector while the interaction mediating between the sectors is described by a non-local form factor. This setup is sufficient to eliminate the divergence due to boost integration, in agreement with intuitive expectations. 
However, in this simplified scenario, the form factor depends on a single invariant, and this will not be enough to control additional phase 
space divergences which may arise when the number decay products is bigger than two. More precisely, we may be left with a 
divergence due to integration over the invariant energies of the decay products within each sector. As we shall see in Section IV, 
this can be remedied by considering fully non-local interactions.  In this case, the rate of decay is finite provided that the interaction is soft 
above some energy scale $\mu$, so that all kinematical invariants constructed out of the decay products are effectively bounded by that scale. 
Again, the rate agrees with the estimate (\ref{Gamma2}). We conclude in Section V.

\section{Local interaction}

Consider a theory with ordinary matter fields $\phi$, and ghost fields $\psi$ in $D$ spacetime dimensions.   
A generic vertex with $n$ ordinary particles and $n'$ ghosts has the form 
\begin{equation}
S_I = \lambda  \int d^D x \  \phi^n(x)\  \psi^{n'}(x). \label{l}
\end{equation}
For simplicity we assume all fields to be scalars. 


To investigate the effect of a finite lifetime of the vacuum, we shall confine the duration of the interaction to a finite time interval, 
\beq
|x^0| \lesssim T. 
\label{xpmz2}
\eeq
This can be implemented by means of a Gaussian regulator,
\begin{equation}
S_I = \lambda  \int d^D x\ e^{-x_0^2/T^2}  \phi^n(x)\  \psi^{n'}(x)\ .\label{rl}
\end{equation}
The interval $\Delta t \sim 2T$ will then play the role of the time elapsed since the creation of the ghost-carrying vacuum, 
{with the hypersurface $t\sim -T$ playing the role of the surface $\Sigma$ on which that vacuum was created.  (Note that boundaries 
between spacetime regions occupied by different vacua are not sharply defined, so the sign "$\sim$" is appropriate here.)}  



From (\ref{l}), the matrix element for production of $n$ ordinary particles of momentum $p_i$ and $m$ ghosts of momentum $k_i$ is given by
\begin{equation}
{\cal M}(0 \to p_1...p_n,k_1,...,k_{n'}) = \lambda   
\int d^D x\ e^{-i(P+K)x}\ e^{-x_0^2/T^2}\label{me},
\end{equation}
where $P \equiv \sum p_i$ and $K\equiv \sum k_i$ are the total momenta of particles and ghosts. 
Integrating over $d^Dx$, we obtain
\begin{equation}
{\cal M} = (2\pi)^D \lambda\ \delta^{D-1}(\bp+\bk)\ \delta_{T}(\Omega)\ , \label{M}
\end{equation}
where $\Omega=P^0+K^0$ and we have introduced the function
\begin{equation} 
\delta_{T}(\Omega) ={T\over 2\sqrt{\pi}}\  e^{-\Omega^2 T^2/4} .
\end{equation}
The latter approaches the standard $\delta(\Omega)$ of energy conservation in the limit 
$T\to \infty$. 

The total probability of decay is given by
\begin{equation}
{\cal P}= \int \prod_{i,j} \tilde d{\bf p}_i \ \tilde d{\bf k}_j\  |{\cal M}|^2,\ \label{prob}
\end{equation}
where we use the notation 
\begin{equation}
\tilde d {\bf p}_i = {d^{D-1}{\bf p}_i\over 2 (2\pi)^{D-1} |p_i^0|}.
\end{equation}
Note that ${\cal M}$ only depends on the total momenta $P$ and $K$ for particles and
ghosts. Hence, it is convenient to factor out the phase space integrals over the individual momenta of the decay products. To do this, we write
\begin{equation}
{\cal P} = \int d^D P \ d^D K \ ds \ d s'\ \delta(P^2-s)\ \delta(K^2- s')\ \rho_\phi(P^2)\ \rho_\psi(K^2)\  |{\cal M}|^2, \label{prob2}
\end{equation}
where we have introduced 
\begin{equation}
\rho_\phi(P^2)\equiv  \int \prod_{i=1}^n \tilde d {\bf p}_i\  \delta^{(D)}\left(P-\sum p_i\right), \quad
\rho_\psi(K^2)\equiv \int \prod_{j=1}^{n'} \ \tilde d {\bf k}_j\ \delta^{(D)}\left(K-\sum k_i\right).\label{spectrald}
 \end{equation}
From (\ref{M}) and (\ref{prob2}), we find the general expression for the inclusive decay rate per unit volume:
\begin{equation}
\Gamma_{total} = {1\over 2T}{d{\cal P} \over dV}= \int ds d s'\ \rho_\phi(s)\rho_\psi(s')\ \Gamma(s, s'),\label{trate}
\end{equation}
where
\begin{equation}
\Gamma(s, s') = \frac{1}{8} (2 \pi)^D \sqrt{\pi}\ \lambda^2 \int {d^{D-1}
\bp\over(\bp^2+s)^{1/2}(\bp^2+ s')^{1/2}} \delta_{T}(\sqrt{2}\ \Omega) 
\label{rate}
\end{equation}
and the parameter $\Omega$ is now expressed in terms of $s, s'$ and
$\bp$,
\beq
\Omega=P^0 + K^0 = (\bp^2+s)^{1/2}-(\bp^2+ s')^{1/2} .
\eeq
Note that $K^0$ is negative because $\psi$ fields are ghosts. 
 
The rate $\Gamma(s,s')$ has to be integrated with the spectral densities $\rho(s)$ in order to obtain 
the total rate (\ref{trate}). 
For the special case of a single stable particle of mass, $n=1$,  the density is 
\begin{equation}
\rho(s) = \delta(m^2-s),  \label{remnant}
\end{equation}
where $m$ is the particle's mass.
For $n>1$, we will be mostly interested in the form of $\rho(s)$ at large values of s, much greater than all particle masses squared.  On dimensional grounds, we can write
\beq
\rho_\phi (s) \sim s^{\frac{D(n-1)}{2}-n}\, \label{rhos}
\eeq
and similarly for $\rho_\psi(s)$.

To connect with standard arguments, let us for a moment consider the limit $T\to \infty$. Then, 
\begin{equation}
\delta_{T}(\sqrt{2}\ \Omega) \to  \delta(\sqrt{2}\ \Omega) = \sqrt{2} \ (\bp^2+s)^{1/2} \delta(s-s').
\label{delta}
\end{equation}
Using (\ref{rate}) and (\ref{trate}), we have
 \begin{equation}
\Gamma_{total} =\frac{1}{8}(2\pi)^{D+\frac{1}{2}} \  \lambda^2 \int\  ds\  \rho_\phi(s)\rho_\psi(s)\ \Gamma(s), 
\label{trate2}
\end{equation}  
where
\begin{equation}
\Gamma(s) = \int {d^D\bp\over(\bp^2+s)^{1/2}}\ . \label{gammas}
\end{equation}
The momentum integral in (\ref{gammas}) leads immediately to the well known divergence associated with boost integration, as in (\ref{Gamma}).


Returning now to finite values of $T$, let us examine the momentum integral $\Gamma(s,s')$ in (\ref{rate}).  
At large values of $|\bp|$, $\bp^2\gg s,s'$, we have $\Omega\approx (s-s')/2|\bp| \ll 1$ and $\delta_T (\sqrt{2}\Omega)\approx const$.  Hence, for $D>2$ the integral is still divergent.  We shall discuss what can be done about this divergence in the following sections.  But before getting to that, let us stop to consider the special case of $D=2$, when the integral (\ref{rate}) is convergent.

In this case, $\Gamma(s,s')$ can be estimated as 
\beq
\Gamma(s,s') \sim \lambda^2 {\rm min}\{|s-s'|^{-1} \ , T/\sqrt{s+s'}\} 
\sim \lambda^2 (|s-s'|+ \sqrt{s+s'}/T)^{-1}.  \label{Gammas2D}
\eeq
To estimate the total rate (\ref{trate}), we note that for $D=2$ it follows from (\ref{rhos}) that  the spectral 
densities are  $\rho_\phi(s) \sim \rho_\psi(s) \sim s^{-1}$ for all values of $n,n'>1$.  Substituting this and (\ref{Gammas2D}) in 
Eq.(\ref{trate}) for the total rate, we obtain
\beq
\Gamma_{total}\sim \lambda^2 \int_{\sim m^2}^\infty\frac{ds}{s}  \int_{\sim m^2}^\infty \frac{ds'}{{s'}}  (|s-s'|+ \sqrt{s+s'}/T)^{-1}
\sim \lambda^2 m^{-2} \ln(mT).  \label{trate2D}
\eeq
Here, the lower bounds on $s$ and $s'$ integrations are set by the particle masses, which we assume for simplicity to have the 
same order of magnitude, $\sim m$.  Note that the total vacuum decay rate (\ref{trate2D}) increases logarithmically with time $T$, 
as one might expect from the logarithmic divergence of the boost integral (\ref{gammas}) in $2D$.  

In the case of physical interest, $D=4$, the vacuum decay rate (\ref{rate}) is divergent even at finite values of $T$.  In the 
following sections we will show that this divergence can be removed if we allow the interaction between the particle and ghost 
sectors to be nonlocal.  In what follows, we focus on the case of $D=4$.

\section{Restricted non-locality}

We shall first consider a simplified model in which interactions are local within the particle and ghost sectors, while the interaction mediating between the sectors is non-local, being described by a Lorentz invariant form factor $g(z)$.  The corresponding interaction Lagrangian, generalizing Eq.(\ref{l}), has the form
\begin{equation}
S_I = \lambda  \int d^4 x \int d^4 z\  \phi^n(x+z)\  g(z)\  \psi^{n'}(x-z)\ .\label{nl}
\end{equation}
A standard local interaction is recovered if we take
 \begin{equation}
 g(z)=\delta^{(4)}(z). \label{local}
 \end{equation}
As before, we restrict the duration of the interaction to a finite time interval, 
\beq
(x^0 \pm z^0)^2 \lesssim T^2, 
\label{xpmz}
\eeq
by means of a Gaussian regulator.  Then the dependence on $x^0$ and $z^0$ factorizes, as $e^{-x_0^2/T^2} e^{-z_0^2/T^2}$. 
On the other hand, it may be useful to formally allow for two different time-scales $T'$ and $T$ in the two factors of the regulator: 
\begin{equation}
S_I = \lambda  \int d^4 x\ e^{-x_0^2/T'^2} \int d^4 z\  \phi^n(x+z)\  g(z)\ e^{-z_0^2/T^2} \psi^m(x-z)\ .\label{rnl}
\end{equation}
In the limit $T'\to \infty$ the interaction is translation invariant, and energy is conserved even if we keep $T$ finite. In what follows, unless otherwise stated, we shall keep both of them finite. The case of our interest will be $T\sim T'$, corresponding to an interaction of finite duration.
 
With a nonlocal interaction (\ref{rnl}) the matrix element (\ref{M}) for particle and ghost production generalizes to
\begin{equation}
{\cal M} = \lambda   \int d^4 z\ g(z)\ e^{-z_0^2/T^2}\  
\int d^4 x\ e^{-iP(x+z)}e^{-iK(x-z)}\ e^{-x_0^2/T'^2} 
= (2\pi)^4 \lambda\ \delta^3(\bp+\bk)\ \delta_{T'}(\Omega)\ W_T(w,\bp), \label{Mnl}
\end{equation}
where
\begin{equation}
W_T(w,\bp)= \int d^4 { z}\ g(z)  e^{-iw{z^0}}e^{2i\bp{\bf z}} e^{-z_0^2/T^2}  \label{W} 
\end{equation}
and we have introduced the notation $w=(\bp^2+s)^{1/2}+(\bp^2+ s')^{1/2} $.
The total decay probability is given by Eq.(\ref{trate}) with
\begin{equation}
\Gamma(s, s') = {2 \pi^{4}\sqrt{\pi}\ \lambda^2} \int {d^3\bp\over(\bp^2+s)^{1/2}(\bp^2+ s')^{1/2}} |W_T|^2\ \delta_{T'}(\sqrt{2}\ \Omega) .
\label{ratenl}
\end{equation}

Expressing the form factor $g$ in terms of its Fourier transform $G$,
\begin{equation}
g(z^2) = (2\pi)^{-4} \int dE \ d^3 {\bf q}\  e^{i (Ez^0- {\bf q z})}\  G(E,{\bf q}),\label{GEq}
\end{equation}
and substituting in (\ref{W}), have
\begin{equation}
W_T = \int_{-\infty}^{\infty} dE\  \delta_T(w-E)\ G(E, 2\bp). \label{fW}
\end{equation}
In general, we can split the Lorentz invariant function $G(E,{\bf q})$ into negative and positive
frequency parts
\begin{equation}
G(E, 2\bp) = \theta(E) \ G^+_{S} + \theta(-E) \ G^-_{S},\label{pnf}
\end{equation}
where, we have introduced the invariant
\begin{equation}
 S \equiv (E/2)^2-\bp^2. 
 \end{equation}
Note that $G^{\pm}_S$ should vanish for $S<0$, otherwise we would have imaginary energies at low 
momenta (corresponding to exponential growth in the integrand in (\ref{GEq}) at large time-like separations).

Substituting (\ref{fW}) in (\ref{trate}), we have
\begin{equation}
\Gamma_{total} ={2 \pi^{4}\sqrt{\pi}\ \lambda^2}\sum_{\sigma,\sigma'=\pm}  \int ds\ ds'\ dS\ dS'\ \rho_\phi(s)\rho_\psi(s')\ \Gamma^{\sigma\sigma'}(s,s',S,S'),
\end{equation}
where
\begin{equation}
\Gamma^{\sigma\sigma'}=  G_S^\sigma \ G_{S'}^{*\sigma'}  
\int {d^3 \bp} {\delta_{T'}(\sqrt{2}\ \Omega)\  \delta_T(w-E_\sigma)\ \delta_T(w-E'_{\sigma'}) 
\over (\bp ^2 +s)^{1/2}(\bp ^2 +s')^{1/2}(\bp ^2 +S)^{1/2}(\bp ^2 +S')^{1/2}}.
\end{equation}
Here, 
\begin{equation}
E_\pm= \pm 2 (\bp^2+S)^{1/2}, \quad E'_\pm= \pm 2 (\bp^2+S')^{1/2}.
\end{equation}
Note that the integral over momenta is manifestly convergent. Further, we note that, unless 
$\sigma=\sigma'=+$, the $\delta_T(w-E_\sigma)$ exponentially switches off the interaction in the limit $T\to \infty$. 

As an illustrative example consider the form factor 
\beq
G^{\pm}_S 
= I_{\pm} \delta(S-s_0).  \label{Gdelta}
\eeq  
For $T\gg s_0^{1/2}$, the $G^-_S$ contribution drops out and the rate simplifies to 
\begin{equation}
\Gamma_{total}\propto \lambda^2 I_+^2 T \int ds\ ds'\ \rho_\phi(s)\rho_\psi(s')  
\int {d^3 \bp} {\delta_{T'}(\sqrt{2}\ \Omega)\ \delta_T(\sqrt{2}(w-E_0))
\over (\bp ^2 +s)^{1/2}(\bp ^2 +s')^{1/2}(\bp ^2 +s_0)}.
\end{equation}
where $E_0= 2 (\bp^2+s_0)^{1/2}$. 
The momentum integral converges, and is dominated by momenta of order
$p_* \sim (s+s')T$. Hence, we can estimate 
\begin{equation}
\Gamma_{total}\sim \lambda^2 I_+^2 T T' \int ds\ ds'\ \rho_\phi(s)\rho_\psi(s')  (s+s')^{-1}. \label{gtes}
\end{equation}

We thus see that the quadratic divergence in the decay rate which occurs in (\ref{Gamma}), due to integration over the total momentum 
$\bf P$ of the decay products, disappears in the present context.  Instead, we obtain a rate that grows quadratically in time, 
in agreement with the heuristic arguments presented in the Introduction.   It can be verified that these conclusions are not restricted to our specific example (\ref{Gdelta}) and apply to a wide class of form factors.

However, we should also consider the integral over the invariant  energies 
$s$ and $s'$ of the decay products. When $s$ and $s'$ are large compared with the masses
of the species involved, the spectral densities are given by (cf. Eq.(\ref{rhos}) with $D=4$)
\begin{equation}
\rho_\phi(s)\sim s^{n-2},\quad \rho_\psi(s')\sim (s')^{n'-2} \label{estimates}
\end{equation}
Hence, for $n+n'>2$ the
integrals over $s$ and $s'$ in (\ref{gtes}) would not converge unless we impose some kind of high energy cut-off.
\footnote{The rate would be finite without any cut-off in the invariant energies if we assumed that $G^+_S=0$, 
but then the two sectors would be disconnected at $T\to \infty$, where $\delta_T(w+E_0)\to 0$.}

Let us compare the present situation with the case discussed by Kaplan and Sundrum \cite{Sundrum}, where the integral over total momentum ${\bf P}$ is regularized by means of a Lorentz breaking cut-off.  They noted that the integral over the invariant energy $s$ of the decay products would also be divergent and pointed out that this divergence could be removed by new (Lorentz invariant) physics which might soften the interaction at energies above 
some scale $s_0$ (which they took to be of the same order as the Lorentz breaking cut-off). 
Here, instead, we find that even if the interaction only carries invariant energy of order $s_0$, we still have a divergence when we 
integrate over the invariant energies of the decay products $s$ and $s'$. This may seem surprising, since we are used to thinking that the invariant
energy of the mediator should be the same as that of the decay products. However, in the present case energy is not conserved because the interaction is switched on and off. As a result, at momenta of order $p_*\gtrsim (s+s')T$, the $\delta_T(\Omega)$ of ``energy conservation" allows $s$ and $s'$ to be very different from each other, and to be much larger than $s_0$.


In the following Section we explore a model where the interaction vertex between particles and ghosts is fully non-local, so that the form factor
depends on more than just one kinematical invariant. We shall also assume that interactions become soft when the values of these invariants exceed the 
scale $\mu$. As we shall see, this may lead to acceptable decay rates in phenomenologically interesting scenarios.

\section{Fully non-local interaction}

Here, we consider non-local vertices where the different legs of matter in a given sector do not necessarily 
meet at the same point.
For simplicity, we restrict attention to the four leg vertex:
\begin{equation}
S_I =\lambda \int d^4 x\ d^4 z\ d^4 y_1\ d^4 y_2\  \phi(x+z+y_1) \phi(x+z-y_1) g(z,y_i) \psi(x-z+y_2)\psi(x-z-y_2). \label{fp}
\end{equation} 
The regulated amplitude is then given by 
\begin{equation}
{\cal M} = (2\pi)^4 \lambda\ \delta^3(\bp+\bk)\ \delta_{T'}(\Omega)\ W_T(w,\bp,p,k), \label{M2}
\end{equation}
where, as usual, $P=p_1+p_2$, $K=k_1+k_2$, $\Omega=P^0+K^0$, $w=P^0-K^0$, 
and we have introduced $p=p_1-p_2$, $k=k_1-k_2$ and
\begin{equation}
W_T(w,\bp,p,k)= \int d^4z \ d^4 y_1\ d^4 y_2\ \ g(z,y_i)  e^{-i(w{z^0}-2\bp{\bf z})} e^{-i(p\cdot y_1+k\cdot y_2)}
 e^{-[(z^0)^2+(y_1^0)^2+(y_2^0)^2]/T^2}.\label{Wnl}
\end{equation}
In terms of the form factor in Fourier space $G$, defined by
\begin{equation}
g(z,y_i)=(2\pi)^{-12} \int d^4q\ d^4 q_p\ d^4 q_k \  \ e^{i (q\cdot z+q_p\cdot y_1+ q_k\cdot y_2)}
G(q, q_p,q_k),
\end{equation}
we have 
\begin{equation}
W_T = \int_{-\infty}^{\infty} dE\ dE_p\ dE_k \ \delta_T(w-E)\ \delta_T(p^0-E_p)\ \delta_T(k^0-E_k)\  
G(q,q_p,q_k), \label{fW3}
\end{equation}
where in the last equation $q=(E,2\bp)$, $q_p=(E_p,{\bf p})$ and $q_k=(E_k,{\bf k})$. 
Introducing the invariants
\begin{eqnarray}
S &=& q^2/4 = E^2/4- \bp^2, \label{mandel}\\
S_p&=&q_p^2= E_p^2 -\ {\bf p}^2, \\
S_k&=&q_k^2= E_k^2 -\ {\bf k}^2,     \label{mandelstam}
\end{eqnarray}
we may decompose $G$ into negative and positive frequencies for each one of the energies.
\begin{equation}
G(q, q_p,q_k) =\sum_{\sigma,\sigma',\sigma''=\pm} \theta(\sigma E)\theta(\sigma' E_p)\theta(\sigma'' E_k)
G^{\sigma\sigma'\sigma''}_{q,q_p,q_k}
\end{equation}
Then, we may write
\begin{equation}
W_T = {1\over 4} \sum_{\sigma,\sigma',\sigma''=\pm} \int dS\ dS_p\ dS_k \ {\delta_T(w-E^\sigma)\ 
\delta_{T}(p^0-E_p^{\sigma'})\ 
\delta_{T}(k^0-E_k^{\sigma''}) \over 
(\bp^2+S)^{1/2}({\bf p}^2+S_p)^{1/2}  ({\bf k}^2+S_k)^{1/2} }\ G^{\sigma\sigma'\sigma''}_{q,q_p,q_k}. 
\label{fW4}
\end{equation}
Note that $G^{\sigma\sigma'\sigma''}_{q,q_p,q_k}$ depends on the subscripts only trough invariants constructed
out of the momenta $q,q_p$ and $q_k$. These are $S,S_p,S_k$ and the products 
\begin{equation}
\pi_p=q\cdot q_p,\quad \pi_k=q\cdot q_k, \quad \pi_{pk}= q_p\cdot q_k.
\end{equation}
The decay rate is given by
\begin{equation}
\Gamma_{total} = 8\pi^4 \sqrt{\pi }\lambda^2 \int \prod_{i,j=1}^2 \tilde d{\bf p}_i \ \tilde d{\bf k}_j\  \delta^3(\bp+\bk)\ \delta_{T'}(\sqrt{2}\ \Omega)\ |W_T|^2.\label{tot}
\end{equation}
We have $(3-1)\times 4-3=5$ powers of momentum from the phase space integrals.
On the other hand, there are 6 powers of momentum in the denominator of $|W_T|^2$. Also, assuming that the non-local interaction becomes soft at a short 
distance scale $\mu^{-1}$, the domain of the integration in (\ref{fW4}) is effectively finite $S,S_p,S_k \lesssim \mu^2$, and we do not expect
divergences in the decay rate. 

Let us define ${p}_*\equiv \mu^2 T$, and assume that $T \gg \mu^{-1}$. 
Then, we can estimate 
\begin{equation}
W_T \approx G(q,q_p,q_k). \quad\quad\quad  
( |{\bf p}_i|,|{\bf k}_i| \lesssim {p}_*)
\end{equation}
Then, assuming that the rate (\ref{tot}) is dominated
by momenta of order ${p}_*$, we have
\begin{equation}
\Gamma_{total} \sim {\pi^4 \sqrt{\pi}  \over 2 (2\pi)^{12}} \lambda^2 T' \int_{{\bf p}_i^2,{\bf k}_i^2\sim {p}_*^2}  
{d^3{\bf p}_1 d^3{\bf p}_2\ d^3{\bf k}_2\ d^3{\bf k}_2\over |p^0_1\ p^0_2\ k^0_1\ k^0_2|}\ \delta^3(\bp+\bk)\ |G(q,q_p,q_k)|^2.\label{tott}
\end{equation}
As mentioned above, the amplitude $G$, depends in principle on the 6 invariants we may construct with $q,q_p$ and $q_k$. Assuming that
the interaction becomes soft at the energy scale $\mu$, we may require $S,S_p,S_k,\pi_p,\pi_k,\pi_{pk} \lesssim \mu^2$. The first three of 
these constraints will restrict the values of the energies $E,E_p,E_k$
to be very close to the modulus of the corresponding spatial momenta. For instance,
\begin{equation}
E = 2\sqrt{\bp^2 + S} = 2|\bp| + O(S/|\bp|) \approx 2|\bp| + O(1/T).
\end{equation}
where in the last step we have assumed $S\lesssim \mu^2$ and $|\bp|\sim {p}_* = \mu^2 T$. Similarly, from $S_p,S_k \lesssim \mu^2$ we have 
\begin{equation}
E_p \approx |{\bf p}| +O(1/T), \quad\quad E_k \approx |{\bf k}| +O(1/T).
\end{equation}
Now, from the condition $\pi_p \lesssim \mu^2$ we have 
\begin{equation}
\pi_p = q\cdot q_p = E\ Ep - 2 \bp {\bf p} \approx  2|\bp||{\bf p}|-2 \bp {\bf p} + O({p}_*/T) \lesssim \mu^2.
\end{equation}
The last constraint implies that $\bp$ and ${\bf p}$ have to be aligned within an angle 
\begin{equation}
\theta_p \lesssim \mu/{p}_*. \label{zp}
\end{equation}
A similar argument can be made for $\pi_k$ and $\pi_{pk}$, leading to additional constraints in the alignment of spatial 
momenta
\begin{equation}
\theta_k, \theta_{pk} \lesssim \mu/{p}_*. \label{zk}
\end{equation}
To satisfy (\ref{zp}) and (\ref{zk}), ${\bf p}$ and ${\bf k}$ must remain within a solid angle 
$\Delta\Omega_p, \Delta\Omega_k \sim \pi \mu^2/{p}_*^2$ of the total momentum $\bp$, effectively suppressing
the phase space available by a factor of $\pi^2 \mu^4/{p}_*^4$ (while the integration over the total momentum $\bf P$ 
spans the full solid angle $4\pi$).  From this, the rate (\ref{tott}) can be estimated as
\begin{equation}
\Gamma_{total} \sim {2\pi^7\sqrt{\pi} \over (2\pi)^{12}}\ \lambda^2 T' p^5_*\ {\mu^4 \over p_*^4} |G|^2 \sim\ 10^{-6} {\mu^{8}\over M_p^4}\ (\mu T)^2, \label{fnlr}
\end{equation}
in agreement with the heuristic estimate (\ref{Gamma2}). In the last step we assume $T\sim T'$ and an interaction of gravitational strength $\lambda|G|\sim \mu^2M_P^{-2}$, where $M_p$ is the Planck mass.  

\section{Discussion}

Vacuum is unstable in the presence of ghosts, and formal calculations give a divergent rate of vacuum decay.  Here we found that the rate may actually 
be finite: the divergent momentum integration may be effectively cut off due to the finite lifetime of the metastable vacuum.
We found, however, that this mechanism can work only if the interaction between the particle and ghost sectors is nonlocal above some characteristic energy scale $\mu$.   
The momentum cutoff $P_{max}$ is then 
\beq
P_{max}(t)\sim \mu^2 t,
\label{Pmax2}
\eeq
and assuming that the ghost-particle interaction is gravitational in nature, the vacuum decay rate is given by Eq.~(\ref{fnlr}), 
\begin{equation}
\Gamma_{total}(t) \sim\ A {\mu^{8}\over M_p^4}\ (\mu t)^2,
\label{Gamma3}
\end{equation}
in agreement with the heuristic argument given in the Introduction.  Here, $t$ is the time elapsed 
since the creation of the vacuum,
and $A$ is a numerical phase space factor, which we estimated to be of order $A\sim 10^{-6}$.

Throughout the paper we assumed for simplicity that the metastable vacuum was created on a flat hypersurface $t=0$ in Minkowski space.  
More realistically, such a vacuum could be formed as a bubble in the course of eternal inflation.  The role of the vacuum creation surface 
would then be played by a hyperbolic 3-surface of constant negative curvature.  For a Minkowski vacuum inside the bubble, we expect that the 
same estimate (\ref{Gamma3}) for the vacuum decay rate 
should apply, with $t$ now being the proper time in the FRW coordinates inside the bubble.  For a de Sitter vacuum with energy density $\rho_v >0$, 
this estimate should still hold for $t\lesssim H_v^{-1}$, where $H_v =(8\pi G\rho_v/3)^{1/2}$ is the Hubble expansion rate in that vacuum.
Ghost production in de Sitter remains an interesting problem for further research.

Is it conceivable that our own vacuum is unstable with respect to production of particles and ghosts?  In such a scenario, the time $t$ 
in Eqs.~(\ref{Pmax2}),(\ref{Gamma3}) should be identified with the present cosmic time, $t_0 \sim 10^{10}$~yrs, and the characteristic energy 
of the particles being produced at present is given by 
\beq
E\sim P_{max}(t_0) \sim 10^{-2} \mu^2_{-3} M_p,
\label{E}
\eeq
where $\mu_{-3}= \mu/10^{-3}~{\rm eV}$.

The total energy density of particles (and ghosts) produced up to the present time is
\beq
\rho_\psi \sim \Gamma(t_0) P_{max}(t_0) t_0 \sim A \frac{\mu^8}{M_p^4}(\mu t_0)^4.
\label{rho}
\eeq
This constitutes a fraction
\beq
\frac{\rho_\psi}{\rho_0} \sim A \left(\frac{\mu}{M_p}\right)^6 (\mu t_0)^6  \sim  A\left(\frac{P_{max}} {M_p}\right)^6 \sim 10^{-12} A\ \mu^{12}_{-3}
\eeq
of the total energy density of the universe, $\rho_0 \sim M_p^2/t_0^2$.  

As we shall see, the cutoff scale $\mu$ must be rather low. This means that to the lowest order in pertrubation theory,
only massless (and very light, $m\lesssim \mu$) particles can be produced.\footnote{Heavy particles, of mass $m\gg \mu$. may  be produced by multiple graviton exchange. In particular, semiclassical pairs of Bondi dipoles may be produced non-perturbatively \cite{EG}. The instantons describing such processes
are Lorentz invariant, and as a result the corresponding decay rates do not contain boost divergences: the rates are finite without the need of 
regulating the duration of interaction (see e.g.  \cite{GSV}).}
Unstable particles will eventually decay into gamma rays, protons, electrons and neutrinos, and possibly into some presently unknown stable particles.  High-energy photons will scatter on the CMB, producing $e^+ e^-$ pairs and cascading down in energy.  The resulting background of cascade photons with energies $\lesssim 100$~GeV is constrained \cite{Berezinsky,Ahlers} by Fermi-LAT measurements \cite{Abdo} to have energy density less than 
\beq
\rho_{cas}^{max}\sim 10^{-6}~ {\rm eV/cm^3} \sim 10^{-9}\rho_0.  
\eeq
Assuming that the cascade radiation accounts for a substantial fraction of the total energy density (\ref{rho}) of the produced particles\footnote{This is a reasonable assumption if the low-energy sector of the theory is described by the Standard Model.  For example, strongly interacting heavy particles will decay producing hadronic cascades, which eventually convert into photons, electrons and neutrinos, plus a smaller number of protons.  The only caveat is that most of the energy could be channelled into some unknown weakly interacting stable particles.}, the corresponding bound on $\mu$ is
\beq
\mu\lesssim (1.8 - 5.6)\cdot 10^{-3}~{\rm eV},
\label{bound1}
\eeq
where the lower value corresponds to $A\sim 1$ and the higher one to $A\sim 10^{-6}$.

On the other hand, we should make sure that the nonlocal modification of gravity we assumed here is consistent 
with the precision short-distance tests of the Newton's law.  As of now, the inverse square law for forces of 
gravitational strength has been tested \cite{Adelberger} down to distances $l\sim 0.1$~mm. This corresponds to 
the energy $\sim 2\cdot 10^{-3}$~eV, and so we should require 
\begin{equation}
\mu \gtrsim 2\cdot 10^{-3} {\rm eV}, 
\end{equation}
which is marginally consistent with (\ref{bound1}).

As we mentioned in the Introduction, a nonlocal interaction generally leads to violations of causality. 
With $\mu\sim 10^{-3}$~eV, such violations should occur on relatively large distance scales, up to 0.1~mm. 
For electroweak strength interactions these violations could manifest themselves in scattering experiments \cite{Wise2}.  
However, for a gravitational strength interaction we seem to be in no danger of running into conflict with observational bounds.

Throughout this paper, we have ignored the case of mixing at the quadratic order between the ordinary matter and the ghost sector. 
In the case of the energy-symmetric scenario of Ref. \cite{Sundrum} such mixing does not occur, at least to the lowest order in 
perturbation theory which we have considered here. Nonetheless, this may be relevant in more general situations such as, 
for instance, the dark energy model considered in Ref. \cite{Antoniadis}. Consideration of these extensions is left for further research.

\subsection*{Acknowledgements}

We are grateful to Gregory Gabadadze for collaboration at early stages of this work and for many
clarifying discussions later on.  We are also grateful to Gia Dvali and Slava Mukhanov for stimulating 
discussions related to the subject of the present paper, and for sharing with us a draft of their
work in progress \cite{GCS}. This work was supported in part by grants AGAUR 2009-SGR-168, 
MEC FPA 2007-66665-C02, MEC FPA 2010-20807-C02-02 and CPAN CSD2007-00042 Consolider-Ingenio 2010 (JG) and 
by the National Science Foundation grant PHY-0855447 (AV).

\end{document}